\documentstyle[graphicx]{mn}
\title{The evolution of core and surface magnetic field in isolated 
        neutron stars}
\author[D.~Konenkov and U.~Geppert]{D.~Konenkov$^{1,2})$ and U.~Geppert$^2)$  \\
     1) A.F.Ioffe Institute of Physics and Technology,
        Politechnicheskaya 26,
	194021 St.Petersburg, Russia \\
        {\it e-mail: dyk@astro.ioffe.rssi.ru} \\
     2) Astrophysikalisches Institut Potsdam,
        An der Sternwarte 16,
        D-14482 Potsdam, Germany \\
        {\it e-mail: urme@aip.de} }

\date{Accepted ...
      Received ...;
      in original form ... }

\def\la{\;\raise0.3ex\hbox{$<$\kern-0.75em\raise-1.1ex\hbox{$\sim$}}\;}
\def\ga{\;\raise0.3ex\hbox{$>$\kern-0.75em\raise-1.1ex\hbox{$\sim$}}\;}

\begin{document}

\maketitle

\begin{abstract}

We apply the model of flux expulsion from the superfluid and
superconductive core  of a neutron star, developed by
Konenkov \& Geppert (2000), both to neutron star models based on different
equations of state and to different initial magnetic field structures. 
When initially the core and the surface magnetic field are of the same order of
magnitude, the rate of flux expulsion from the core is almost independent of 
the equation of state, and the evolution of the surface field decouples from
the core field evolution with increasing stiffness.
When the surface field is initially much stronger than the core field, the
magnetic and rotational evolution resembles to those of a neutron star with
a purely crustal field configuration; the only difference is the occurence of
a residual field.
In case of an initially submerged field significant differences from the
standard
evolution occur only during the early period of neutron star's life, until
the field has been rediffused to the surface. The reminder of the episode of
submergence is a correlation of the residual field strength with the
submergence depth of the initial field. We discuss the effect of the 
rediffusion of the magnetic field on to the difference between the real and 
the active age of young pulsars and on their braking indices. Finally, we 
estimate the shear stresses built
up by the moving fluxoids at the crust--core interface and show that 
preferentially in neutron stars with a soft equation of state these stresses 
may cause crust cracking.

\vspace{0.4cm}

\noindent {\bf Key words}: magnetic fields - stars: neutron - pulsars: general
- stars: evolution 

\end{abstract}

\section{Introduction}
Since the process of the generation of neutron star (NS) magnetic fields (MFs) 
and, hence, their initial structure, strength and localization is still under
discussion, we here intend to investigate the  evolution of NS MFs which 
penetrate the entire star. Though there are arguments that the NS MF may be 
confined to the crustal layer, where it has been generated by thermoelectric 
effects during the early hot period of the NS's life when the temperature 
gradients in the crust are immense (Blandford, Applegate \& Hernquist 1983; 
Urpin,  Levshakov \& Yakovlev 1986; Wiebicke \& Geppert 1996), one has also 
to consider the possibility that the NS MF permeates both the crust and the 
core of the star. This could be the field structure from the very beginning 
of the NS's life, either according to the simple model of flux conservation 
during 
the collapse or due to a very efficient dynamo action in the core of the 
proto--NS as described by Thompson \& Duncan (1993).\\
\noindent In a recent paper (Konenkov \& Geppert 2000, hereafter KG00) we
considered the flux expulsion from the superfluid core of a NS. Since the
transition from the normal to the superfluid state of the matter in the
cores of NSs may occur rather early after the NS's birth (Page 1998), we 
assumed
that the protons in the core form a superconductor of type II
(Baym, Pethick \& Pines 1969) and the magnetic flux is concentrated in
an array of proton flux tubes (fluxoids). There are several forces acting
on to the fluxoids (for details see Sec. 2), driving them outward into
the normal conductive crust of a NS, where the magnetic flux may decay 
ohmically.
Considering an initial field structure where the MF at the surface
and in the core has the same strength, we showed
that the field evolution in the crust leads to a deceleration of the flux
expulsion which is the more pronounced the stronger the initial surface MF
is. It turned out, that the main force, which is responsible for the flux
expulsion from the core, is the buoyancy force.\\
\noindent  We have also confirmed the
result of Ding, Cheng \& Chau 1993 (hereafter DCC), that not the entire
magnetic flux is expelled from the core, but a certain part of it remains
there for eternity. This results in a nonvanishing residual field $B_{res}$,
which can be in the range of $10^7-10^{10}$ G, depending on the model
parameters.\\
\noindent It is well known that the equation of state (EOS), which describes
the state of matter in the core region of the NS, has a crucial influence on
the crustal MF decay in isolated NSs (Urpin \& Konenkov 1997; Page, Geppert \&
Zannias 2000). The EOS determines the compactness of the NS and, hence, the
scale length of the MF in the crust, as well as the cooling history of the NS.
While a softer EOS causes an smaller scale which, in turn, leads to a more
rapid field decay, a softening of the EOS decelerates also the cooling.
In a warmer crust the MF will be
dissipated faster too.  Since the EOS influences also - via the moment of
inertia - the spin--down which is one important process for the core flux
expulsion, we will consider in this paper the field evolution for NSs
modelled with three different EOS,
covering the whole range from very stiff to soft ones.\\
\noindent However, the assumption that the field strengths at the NS's surface 
 and in its core are initially of the same order of magnitude is not the 
only conceivable one. There are good reasons to assume that the field
strengths at these positions differ considerably. This could be, e.g., a 
consequence of a field amplification during the collapse by flux conservation 
which results in a flux permeating the whole star with, say, $10^9$ G and a 
very efficient field generation after the formation of the NS by the 
thermoelectric instability in the crustal layers, producing there, say, 
$10^{12}$ G. Thus, we want to investigate here the effect of different initial 
ratios $B_{p0}/B_{c0} \gg 1$, that is, initial magnetic configurations, for
which the crust contains the bulk of the magnetic flux and only a tiny 
fraction of it is anchored in the core.\\
\noindent Another process which determines the initial MF structure in the 
new--born NS is the post core--collaps accretion of fall--back matter after 
the supernova explosion. If this accretion is hypercritical the ram pressure 
overwhelms the pressure of a possibly existent MF and submerges it. The depth 
of submergence depends on the total amount of accreted matter and on the EOS 
of the NS matter (for details see Geppert, Page \& Zannias 1999). Given the 
parameters as estimated for SN 1987A (Chevalier 1989), a MF generated by dynamo 
action
in the proto--NS or amplified simply by flux conservation during the collaps 
would be submerged down to the crust--core boundary or even into the core. 
The rediffusion of that submerged MF would last at least $10^8$ years, thus, 
the NS born in SN 1987A will not appear as a pulsar in the near future, 
unless field generation processes in the crust will transform thermal into 
magnetic energy relatively fast. However, also fall--back much weaker than 
observed in SN 1987A will cause a certain submergence and, hence, a
delayed switching on of pulsars (Muslimov \& Page 1995).  
The discrepancy between the real age ($1.7 \cdot 10^5$ years) and active age
($1.6 \cdot 10^4$ years) of the pulsar B1757-24 (Gaensler \& Frail 2000)
can be explained at least partly by the assumption that the MF of this
pulsar was submerged and rediffused during $\sim 10^5$ years (see Sec. 3).
Generally, if there was initially some field 
generation in the whole NS, the post--supernova accretion onto the new born 
NS will lead to a field structure with $B_{p0}/B_{c0} \ll 1$, i.e. to a field
structure having a much weaker MF strength at the surface than in the core. 
It is our aim to 
consider here the effect of different submergence depths onto the flux 
expulsion from the core and its consequences for the early NS evolution.\\
\noindent The paper is organized as follows. In Section 2 we give a short 
description of the model. It will be based on the  
model of the interplay of forces that determine the flux expulsion from the 
NS's core presented in detail in KG00. The numerical results for the 
different initial field structures are presented in Section 3 and Section 4 
is devoted to the discussion and conclusions.\\

\section{Description of the model}

We calculate the velocity of the flux carrying proton tubes (fluxoids) using 
the model described in detail in KG00. Namely, fluxoids are supposed not to be 
rigidly tied with the neutron vortices (Srinivasan et al. 1990, 
Jahan-Miri \& Bhattacharya 1994), but to move under the common action of the 
buoyancy force, the drag force and the force exerted by the neutron vortices.\\
\noindent The superfluid core participates in the rotation of the NS by 
forming an array of quantized  neutron vortices.
The radial velocity of vortices at the crust-core boundary is determined by
\begin{equation}
v_n=-\frac{R_c \dot{\Omega}_s} {2\Omega_s},
\end{equation}
where $\Omega_s$ is the averaged angular velocity of the core superfluid, 
$R_c$ is the radius of the NS core. In this paper we consider the
evolution of isolated NSs. We assume, that the rotational evolution is 
determined by energy losses due to magneto--dipole radiation:

\begin{equation}
P\dot{P}=\frac{2 B_p(t)^2 R^6} {3 I c^3},
\end{equation}
where $B_p$ is the surface field strength at the magnetic pole, $R$ is 
the radius of the NS, $I$ its moment of inertia, $c$ is the speed of light and
the magnetic axis is perpendicular to the rotational one.

\noindent The buoyancy force, acting on the unit length of a fluxoid, is given 
by (Muslimov, Tsygan 1985):
\begin{equation}
f_b = \left(\frac{\Phi_0}{4\pi \lambda}\right)^2
        \frac{1}{R_c}\ln\left (\frac{\lambda}{\xi}\right) ,
\label{fb_eq}
\end{equation}
where $\Phi_0=2 \times 10^{-7}$ G cm$^2$ is the quantum of the magnetic flux, 
$\lambda$ is the London 
penetration depth, and $\xi$ is the proton coherence length, which is less than
$\lambda/\sqrt{2}$ for a
superconductor of type II. The buoyancy force is
always positive, i.e. directed outward.

\noindent The drag force per unit length of a fluxoid, resulting from the 
interaction 
of the normal electrons in the core with the magnetic field of a fluxoid 
(Harvey, Ruderman, Shaham 1986, Jones 1987), is given by
\begin{equation}
f_v = -\frac{3\pi}{64}\frac{n_e e^2 \Phi_0^2}{E_F \lambda}\frac{v_p}{c},
\label{fv_eq}
\end{equation}
where $n_e$ is the number density of electrons in the core, which is assumed 
to be 
about 5\% of the number density of neutrons, $e$ is the elementary
charge,  $E_F$ the Fermi energy of the
electrons, and $v_p$ is the velocity of fluxoid. For the 
Fermi energy and electron number density we take the values determined by the 
density at the crust--core interface, $\rho_c=2 \cdot 10^{14}$g/cm$^3$. 
When $v_p>0$ (fluxoids are moving outward), one has $f_v<0$, i.e. $f_v$ 
hampers the expulsion of the flux from the core of NS.

\noindent The force, acting on the unit length of a fluxoid exerted by 
the neutron vortices, is given by (DCC)
\begin{equation}
f_n = \frac{2 \Phi_0 \rho r
\Omega_s (t) \omega (t)}{B_c(t)},
\label{fn_eq}
\end{equation}
where $\omega=\Omega_s-\Omega_c$ is the lag between the rotational velocity of 
the core superfluid, $\Omega_s$, and the rotational velocity of both the crust 
and the charged component of the core, $\Omega_c$, while $B_c$ denotes the
averaged core magnetic field. The observed spin period $P$ of a NS is related 
to $\Omega_c$ by $P=2 \pi /\Omega_c$. The maximum lag $\omega_{cr}$, which can 
be sustained by the pinning force, defines also the maximum force, which can 
be exerted by the vortices on to the fluxoids (see DCC and Jahan--Miri 2000 
for a more detailed discussion).
According to DCC,
\begin{eqnarray}
\omega_{cr}=8.7 \times 10^{-2} x_p \alpha_g r_6^{-1} \left( \frac{m_p - m_p^*}
{m_p} \right) \left( \frac{m_p^*}{m_p} \right)  ^{-1/2} \times \nonumber \\
          \times B_{c12}^{1/2} \ln \left( \frac{\lambda} {\xi} \right)
          \sin(2 \chi) {\rm \,\,\, rad \,\, s} ^{-1},         
\end{eqnarray}
where $x_p$ is the fractional concentration of protons, $\alpha_g$ is a 
numerical factor of order of the
unity, $r_6$ is the distance from the NS 
rotational axis in $10^6$ 
cm, $m_p$ is the mass of the proton, $m_p^*$ its effective mass, $B_{c12}$ is 
the core magnetic field strength in units of $10^{12}$ G, and $\chi$ is 
the angle between rotational and magnetic axis. We assume hereafter 
$x_p=0.025$, $\alpha_g=1$ and $m_p^*=0.8 m_p$

\noindent  Vortices can either move outward faster than fluxoids (forward
creeping, $\omega = \omega_{cr}$), or the velocities of both kinds of flux 
tubes can
coincide (comoving,$-\omega_{cr}<\omega < \omega_{cr}$ ), or neutron vortices
can move slower than fluxoids (reverse creeping, $\omega=-\omega_{cr}$).
The vortex acting force can be both positive or negative, depending on the 
sign of $\omega$, i.e. $f_n$ can either promote the expulsion or impede it.

\noindent The sum of the powers of the forces, acting onto the fluxoids in 
the core, is equal to the Poynting flux through the surface of the NS core:
\begin{equation}
\sum_{fluxoids}\int ({f}_b+{f}_n+{f}_v){v}_p{{\rm d}l}=
-\frac{c}{4\pi}\int\limits_{S_{core}}\left[\vec{E} \times \vec{B} \right] 
\cdot d\vec{S}_{core},
\label{main_eq}
\end{equation}
where the summation runs over all fluxoids. The integral on the l.h.s. of 
(\ref{main_eq})
is taken over the length of each fluxoid, while the integral on the r.h.s. of 
(\ref{main_eq}) is performed over the 
surface of the core; the  normal vector of this surface is directed into the 
core. We consider the evolution of a poloidal field, which is supposed to be
dipolar outside the NS. Introducing a vector potential $A=(0,0,A_{\phi})$ 
and the Stokes stream function $A_{\phi}=S(r,t) \sin(\theta)/r
=B_{p0} R^2 s(r,t) \sin(\theta)/r$,  where 
$r$, $\theta$, $\phi$ are spherical coordinates, $s(r,t)$ the normalized
stream function, and $B_{p0}$ the initial magnetic field strength at the 
magnetic pole, one can express the field components as 
\begin{equation}
B_{r} = B_{p0} R^2 \frac{s}{r^2} \cos \theta,
\,\,\, 
B_{\theta} = - \frac{B_{p0} R^2} {2} \frac{\sin \theta}{r} \cdot 
                         \frac{\partial s}{\partial r}. 
\end{equation}

\noindent By means of the variable $s$ the integral on the r.h.s. of the 
equation (\ref{main_eq}) 
can be rewritten as:
\begin{equation}
\frac{c}{4\pi}\int\limits_{S_{core}}\left[\vec{E} \times \vec{B} \right]
                           \cdot             d\vec{S}_{core} = 
\frac{ B_{p0}^2 R_c^2}{6} \frac{\partial s(R_c,t)} {\partial t}
\frac{\partial s(R_c,t)} {\partial r}.
\end{equation}

\noindent Assuming the MF in the core to be uniform, one can introduce the
total core forces
\begin{equation}
F_{n,b,v}=f_{n,b,v}\cdot 4R_c/3 \cdot N_p,
\label{F_eq}
\end{equation}
where $4 R_c/3$ is the mean length of the fluxoid and $N_p=
\pi R_c^2 B_c / \Phi_0$ is the number of fluxoids. It is worth mentioning 
that $F_n \propto B_c^{1/2}$ both in the forward 
and in the reverse creeping regimes, but $F_b, F_v \propto B_c$.

\noindent One can now introduce the crustal force
\begin{equation}
F_{crust}=
       \frac{B_{p0}^2 R^4}{6} \frac{1}{v_p}
       \frac{\partial s(R_c,t)} {\partial t}
       \frac{\partial s(R_c,t)} {\partial r}. 
\label{F_c_eq}     
\end{equation}
When $\partial s(R_c,t) / \partial r$ is positive and 
$\partial s(R_c,t) / \partial t$ is negative the crustal force is negative, 
i.e. it prevents the roots of the fluxoids to move towards the magnetic 
equator. The crustal force is proportional to $B_c^2$.

\noindent Now one can rewrite equation (\ref{main_eq}) as:
\begin{equation}
F_n+F_b+F_v(v_p)+F_{crust}(v_p)=0,
\label{forces_eq}
\end{equation}
which determines the velocity of fluxoid $v_p$.

\noindent The equation of the balance of electromagnetic energy in the 
region outside NS core 
reads:
\begin{equation}
\int\limits_{V_{\rm crust}}\frac{j^2}{\sigma}\,{d}V + 
     \frac{d}{dt}\int\limits_V \frac{B^2}{8\pi}\,{d}V +
     \frac{c}{4\pi}\int\limits_{S_{core}}\left[\vec{E} \times \vec{B} \right]
     \cdot  d\vec{S}_{core}=0.
\end{equation}
The first integral is restricted to the crust, the second has to be taken
over all space excluding NS's core, and the third is a surface integral 
over  the core. Therefore, the crustal force determined by the equation 
(\ref{F_c_eq}) coincides with the crustal force determined by the
equation (8) from KG00, however, a smoother numerical procedure can is 
achieved with the recent version. \\

\begin{figure*}
\centering\includegraphics[]{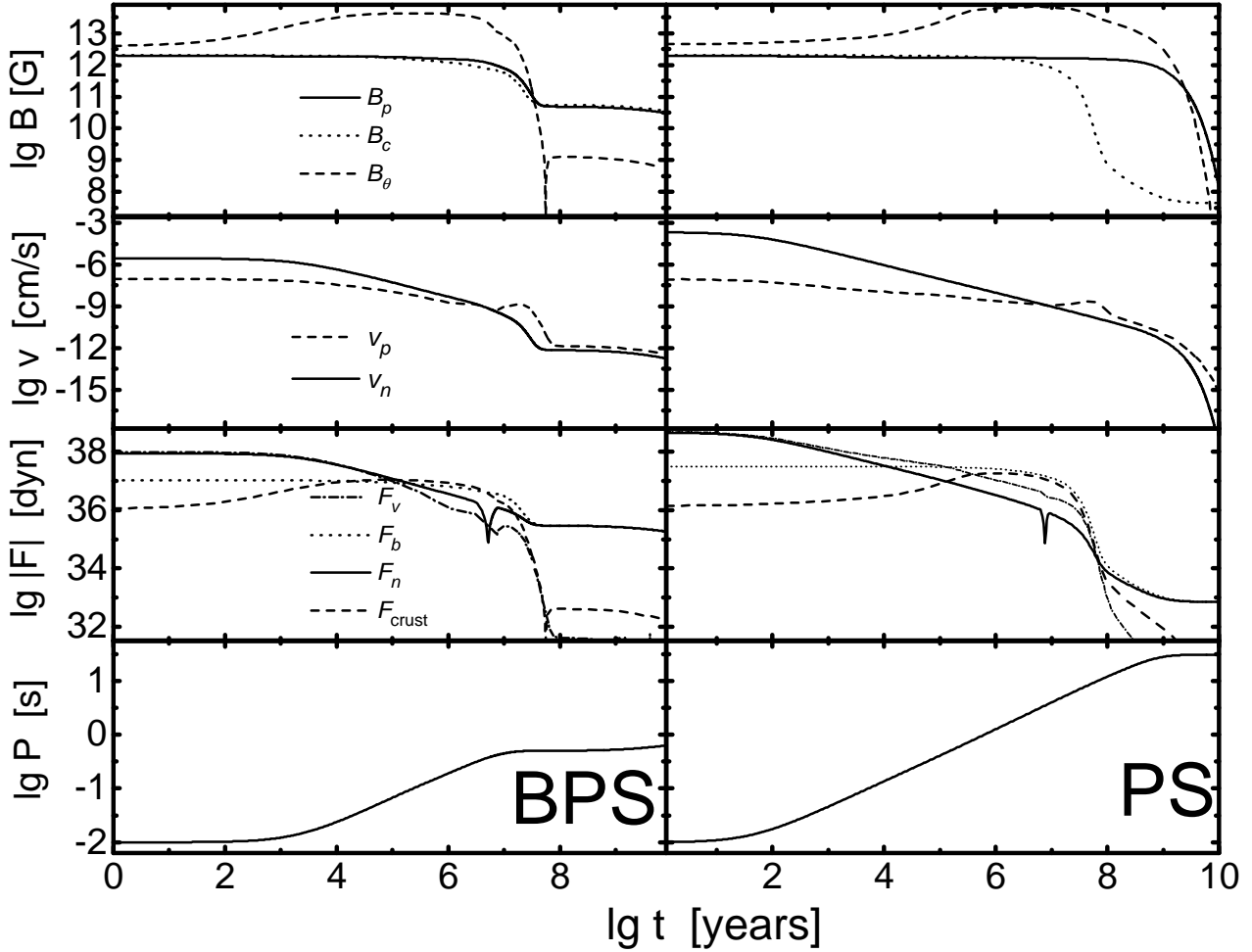}
\caption{The evolution of the magnetic field strengths (surface field at the
magnetic pole $B_p$, core field $B_c$, and $\theta$-component of the field
at the magnetic equator at $R=R_c$), velocities of vortices and fluxoids,
forces and rotational period of a NS, based on BPS (left column) and PS
(right column) EOS. Parameters: $Q=0.1$, $P_0=0.01$, $B_{c0}=B_{p0}=2 \cdot
10^{12}$ G.
}
\end{figure*}

\noindent The evolution of the magnetic field in the solid crust is governed 
by the induction equation without a convective term:
\begin{equation}
\frac {\partial \vec{B}} {\partial t} =
  -\frac{c^{2}}{4 \pi} \nabla \times 
    \left( \frac{1}{\sigma} \nabla \times \vec{B} \right).
\end{equation}
It can be rewritten in terms of the Stokes stream 
function as:
\begin{equation}
\frac{\partial s}{\partial t} = \frac{c^2}{4\pi\sigma}
                   \left(\frac{\partial^2 s} {\partial r^2} 
                    - \frac{2s}{r^2}\right),
\end{equation}
and should be supplemented by the boundary conditions at the surface ($r=R$)
and at the crust-core boundary ($r=R_c$). While the outer boundary condition
is given by $R\frac{\partial s}{\partial r} = -s$ at $r=R$,
the inner one is time-dependent and determined by equation (\ref{forces_eq}).

\noindent The conductivity in the regions of interest in the crust is 
determined by the scattering of electrons at impurities and phonons. The phonon 
conductivity is dependent on the temperature, and, hence, on the cooling 
history of the NS. The impurity conductivity is 
temperature independent and is determined by the concentration of the 
impurities $Q$. The relaxation times for both electron scattering processes
depend strongly on the density and on the chemical composition of the crust.
For the phonon conductivity we use the numerical data given by Itoh et al. 
1993, for the impurity conductivity we apply the analytical expression 
derived by Yakovlev \& Urpin 1980.\\
\noindent The evolution of the magnetic field in the superconducting core 
is driven only by the motions of the fluxoids:
\begin{equation}
\frac {\partial \vec{B_c}} {\partial t} = \nabla \times 
                \left( \vec{v}_p \times \vec{B_c} \right).
\end{equation}
This equation can be rewritten in terms of the normalized stream function 
$s$ by:
\begin{equation}
\frac{\partial s}{\partial t} = -v_p \, \frac{\partial s}{\partial r}.
\label{simpl_conv}
\end{equation}
The assumption of the homogeneity of the magnetic field in the core
leads to the following ansatz for the fluxoid velocity
\begin{equation}
v_p(r,t)=\alpha(t)r,
\label{vel_flux}
\end{equation}
and for $s(r,t)$ in the core
\begin{equation}
s(r,t)=\frac{B_c(t)r^2}{B_{p0}R^2}.
\label{s_core}
\end{equation}
With this, the solution of equation (\ref{simpl_conv}) describes the 
evolution of the core MF at the crust--core boundary  as
\begin{equation}
B_c(t)=B_{c0} \exp \left(-2\int_0^t \alpha(t') dt' \right).
\label{B_core}
\end{equation}
Note, that in equations\ (15) and (17) of KG00 there are two misprints. However,
the numerical computation was based on the correct equations. We use the 
procedure described in KG00 to find the velocity of the
fluxoids and, consequently, the magnetic and rotational evolution of NS.\\

\section{Numerical results}

Although the induction equation is linear, the problem, as shown in Section 2,
is a nonlinear one in terms of the magnetic field, and has a time and field 
dependent boundary condition at the crust--core interface. Additionally, 
equation\ (\ref{forces_eq}) which determines the flux expulsion velocity is
nonlinear in $v_p$ too. For the calculations  described below we use the 
standard cooling scenario considered by Van Riper (1991) for the neutron star 
model based on a stiff (Pandharipande \& Smith 1975), medium 
(Friedman--Pandharipande 1981) and a soft (Baym, Pethick \& 
Sutherland 1971) EOS. The cooling histories for those EOS are taken 
from Van Riper (1991).\\ 
The main influence of the choice of the EOS on to the MF evolution 
comes via the different density profiles which determine the scale lengths
of the MF in the corresponding NS models.\\
In all our model calculations we apply the initial rotational period 
$P_0=0.01$ s, a value which is generally accepted for new--born NSs.

\subsection{The effect of different EOS}
In order to extract the effects of different EOS on to the flux expulsion
from the NS's core we apply for all models the initial surface and core
magnetic field strengths
$B_{p0} = B_{c0} = 2 \cdot 10^{12}$ G, an impurity parameter $Q=0.1$ and the
standard cooling scenario.\\
\noindent We consider three qualitatively different EOS and use the
corresponding density profiles calculated for a $1.4 M_{\odot}$ NS (for details
see Urpin \& Konenkov (1997) and references therein). The soft EOS is
represented by the BPS model with a radius of $7.35$km and a crust thickness of
$310$m. The corresponding values for the medium (FP model) and stiff (PS
model) EOS are $10.61$km , $940$m, and $15.98$km, $4200$m, respectively. 
The effect of increasing stiffness of the EOS describing 
the state of the core matter is at least threefold: i) since
$P \dot P \propto B_p^2\,R^4/M$ the increasing radius of the NS $R$ leads to 
a more efficient spin--down for a given surface MF $B_p$ and NS mass $M$; 
ii) the larger scale of the MF and iii) the faster cooling of the NS cause 
a deceleration of the crustal field decay.\\
\noindent In Figure\ 1 we show the evolution (from top to bottom) of the 
MF, velocities of vortices and fluxoids, forces and rotational periods 
of NS models based on a very soft (BPS) and a very stiff (PS) EOS.
As in the case of the medium FP EOS (see KG00), at the beginning the dominating
expulsive force is $F_n$. This is because in the core of the magnetized and
rapidly spinning NS there exists a large number of fastly outward moving neutron
vortices, which act on the fluxoids. As shown in the third panel of
Figure\ 1, during that relatively short early period of $t < 10^4 - 10^5$ 
years, $F_n$ is balanced by $F_v$, while the other forces in 
eq.(\ref{forces_eq}) are negligible. However, because 
of the preceding NS spindown, the number of vortices in the core 
(and, hence, $F_n$) decreases, and $F_b$ becomes the main force, which drives
the fluxoids outward. In KG00 we have shown, that in case of an intermediate 
EOS, the timescale of the expulsion of the magnetic field from the core 
$\tau_e$ is determined by the balance of $F_b$ and $F_{crust}$. One
can see that, although $F_b$ and $F_{crust}$ are still the dominant
forces governing the expulsion timescale both for NSs with stiff and soft EOS,
in the case of the BPS model one can not neglect $F_n$ in the balance of 
forces. NSs with softer EOS spin down slower than NSs with stiffer EOS. 
In $10^6$ years a NS with the soft EOS spins down to only 0.19 s, while a NS 
with the stiff EOS spins down up to 1.26 s. Thus, $F_n$, which is proportional 
to the number of vortices, i.e. inversely
proportional to the spin period of the NS, plays a more important role in
the dynamics of fluxoids for NSs with a soft EOS.
The evolution of their surface MF follows closely the core field decrease,
because the diffusion timescale of the crustal currents is smaller than 
the expulsion timescale of the core field.
The expulsion of the flux in case of the BPS model stops at about
$3.5 \cdot 10^7$ years, leaving behind a residual field
$B_{res} \approx 5 \cdot 10^{10}$ G, since then the vortex acting force
balances $F_b$ and prevents further expulsion. Therefore, the slight
spindown results in a relatively short residual spin period 
$P_f \approx 0.5$ s and a high strength of the residual MF.\\
\noindent In the case of a NS model with a stiff EOS, the expulsion timescale
appears to be much shorter than the dissipation timescale of the field in the
thick crust. When the MF in the core has been reduced by $\sim 4$ orders of
magnitude (after $\sim 2 \cdot 10^8$ years), the surface field is still
$\sim 10^{12}$ G. Thus, the evolution of the surface MF in NSs with stiff
EOS is insensitive to the details of the physics of the superfluid core
(but, of course, is sensitive indirectly via the cooling history of NS which 
is extremely sensitive to the occurrence of superfluidity). 
In order to check the sensitivity of the results with respect to another 
{\it a priori} unknown parameter of our models, we varied $Q$ in the range 
$0.01-1$ and found, that the surface MF begins to decay for any $Q$ when the 
core flux is almost completely expelled.\\
\noindent Since the NS with stiffening of the EOS spins down more effectively
(because of the larger radius and the longer decay--timescale of the
surface MF), at the end of the evolution a smaller number of neutron vortices 
remains in the core which can hold there only a smaller number of fluxoids 
than in case of a softer EOS NS model. Thus, the residual field becomes 
weaker with stiffening of the EOS.\\

\subsection{The case of $B_{c0}<<B_{p0}$}

\begin{figure}
\centering\includegraphics[width=8cm]{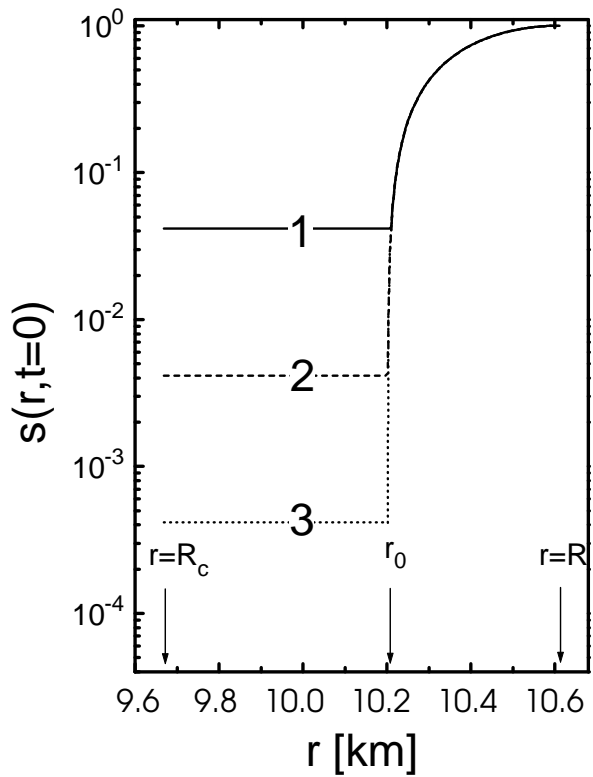}
\caption{The initial profiles of the normalized stream function $s(r,t=0)$
in the crust ($R_c<r<R$).
Lines marked by 1, 2, 3 correspond to $B_{c0}=10^{11}, 10^{10}$ and
$10^9$ G, respectively. The $r=r_0$ corresponds to the density $\rho=\rho_0$.
}
\end{figure}

\begin{figure*}
\centering\includegraphics[]{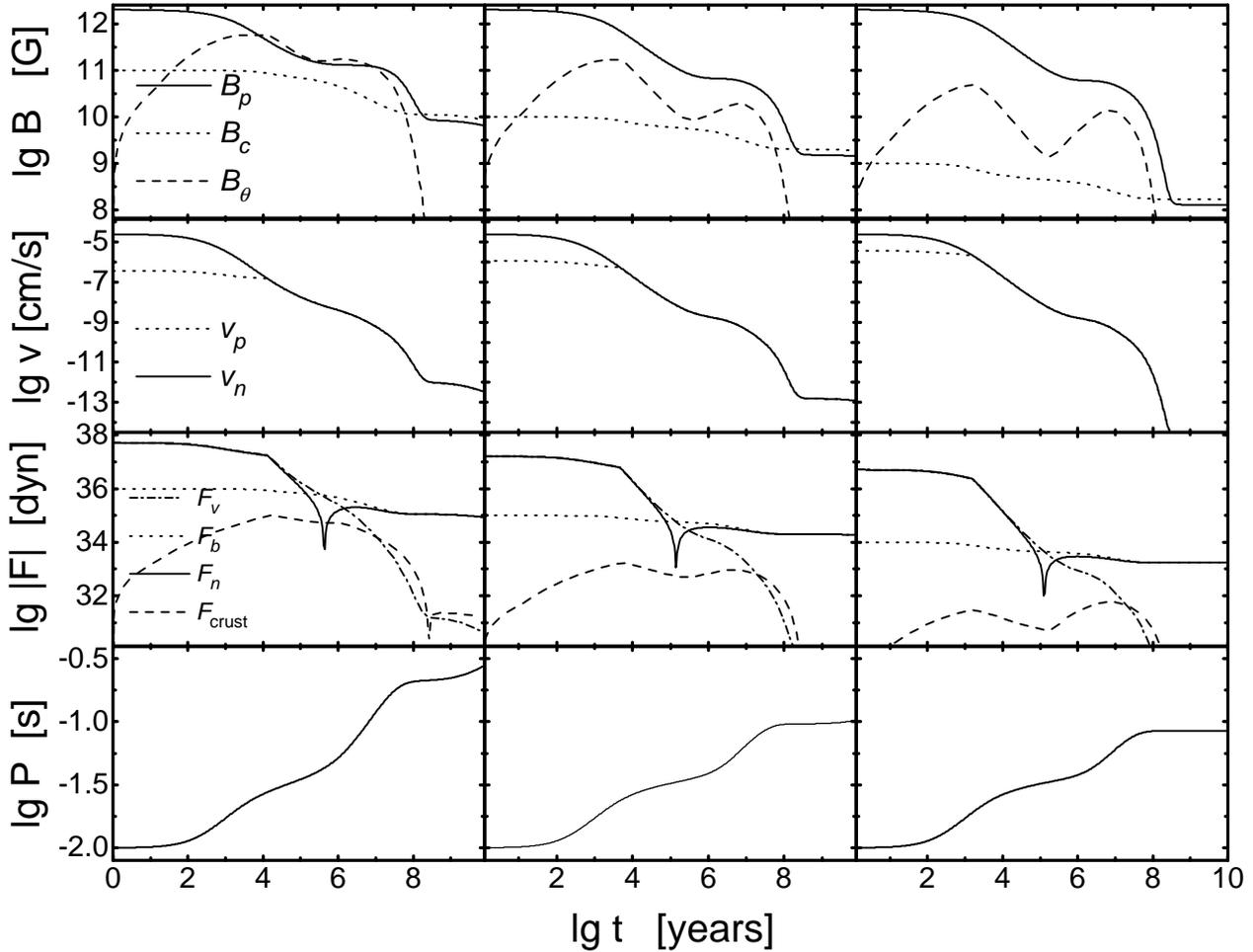}
\caption{The evolution of the magnetic field strengths (surface field at the
magnetic pole $B_p$, core field $B_c$, and $\theta$-component of the field
at the magnetic equator at the $R=R_c$), velocities of vortices and fluxoids,
forces and rotational period of a NS, based on the FP EOS, for $B_{p0}=2
\cdot 10^{12}$ G, $B_{c0}=10^{11}$ G (left column), $10^{10}$ G (middle column) 
and $10^9$ G (right column), for the initial $s$--profiles from Figure 2.
}
\end{figure*}

\begin{figure*}
\centering\includegraphics[]{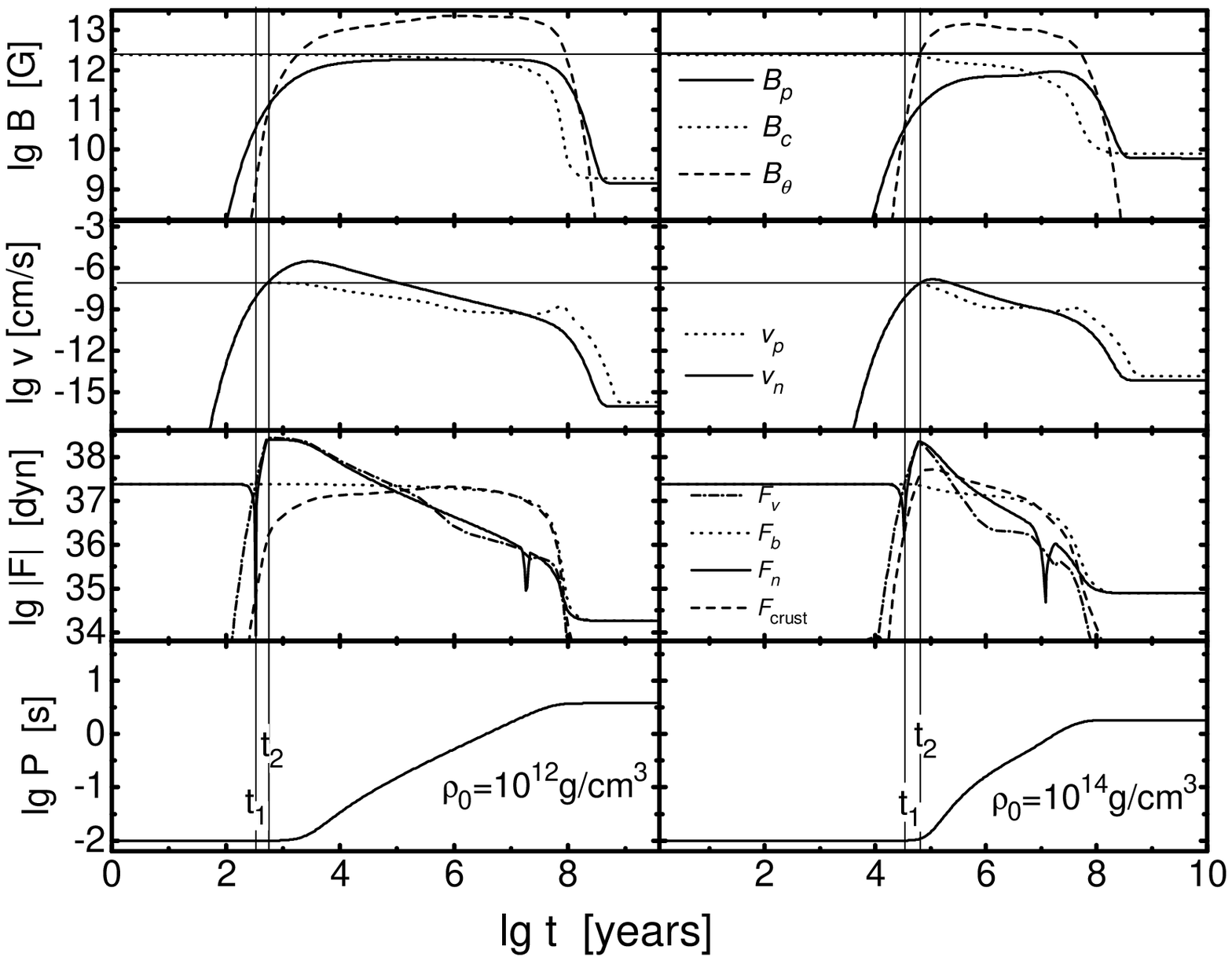}
\caption{The same as in Fig.~1 for the initially submerged field.}
\end{figure*}

\noindent There exist some observational and theoretical evidences that the
initial core field might be weaker than the surface one. Thus, Chau, 
Cheng \& Ding (1992) concluded from an analysis of glitch observations in the 
Vela pulsar that the MF at the crust--core boundary is in the order of 
$10^8 - 10^9$ G while the surface MF is about three to four orders of 
magnitude larger. Such a difference in the field strengths at the surface and 
at the crust--core boundary could be produced by the 
thermoelectric instability (Blandford, Applegate \& Hernquist 1983, 
Urpin,  Levshakov \& Yakovlev 1986, Wiebicke \& Geppert 1996), which may
transform thermal into magnetic energy effectively  during the very early 
period of NS's life in its outermost crustal layers.\\
\noindent In Figure\ 2 we show the initial field configurations in terms of 
the Stoke's stream function $s(r,t=0)$ for $B_{p0}=2\cdot 10^{12}$ G and
$B_{c0}=10^9, 10^{10},$ and $ 10^{11}$ G. 
The calculations have been performed for an initial period $P_0=0.01$ s, an
impurity parameter $Q=0.1$ and an initial density 
where the crustal field matches the inner one, $\rho_0 = 10^{13}$g cm$^{-3}$. 
In order to show clearly the effect of different  $B_{p0}/B_{c0}$, we will not 
vary $Q$ and $\rho_0$ as well as present the results for the medium FP EOS and
standard cooling only.\\
\noindent We show in Figure\ 3 the evolution of the field components,
of the forces acting on to the fluxoids in the core, of the velocities of both 
the fluxoids $v_p$ and of the neutron vortices $v_n$, and the rotational 
evolution in terms of the rotational period $P(t)$.\\
\noindent Since $F_{crust} \propto B_c^2$, $F_b, F_v \propto B_c$, but $F_n 
\propto B_c^{1/2}$, a relative decrease of the field strength in the core
reduces $F_{crust}$, $F_b$ and $F_v$ much faster than $F_n$, thereby 
enforcing the role the vortices play for the dynamics of the fluxoids for
the weaker $B_c$. As seen from the third panels of Figure\ 3 apart from the 
early stages of evolution ($t = 10^3...10^4$ years) the fluxoids are 
completely in the {\rm comoving} regime, and their motion is totally 
determined by those of the neutron vortices, and the velocity of fluxoids at 
$t=0$ is the greater, the smaller $B_{c0}$ is. Consequently, a MF configuration 
with $B_{c0} << B_{p0}$ results in a  ``spin--down induced'' flux expulsion as 
described by Jahan-Miri \& Bhattacharya (1994) for accreting NSs, and the 
influence of the crustal force on the dynamics of fluxoids is negligible.\\
\noindent The decay of the surface MF almost fully coincides with that in case 
of the purely crustal magnetic configurations (Urpin \& Konenkov 1997, Page 
et al. 2000). Only in the case of $B_{c0}=10^{11}$ G, the value of $B_p$ during 
the plateau phase at $t=10^5-10^7$ years is slightly higher than in the case of 
$B_{c0}=10^{10}$ or even $10^{9}$ G. However, the most remarkable difference
is that the surface MF decays not down to zero, but to a nonzero residual
value. Since $v_p=v_n$ after $t>10^4$ years the ceasing of the spin--down of the
NS, when $B_p$ has been decayed down to the weak $B_c$, results in a ceasing
of the further field expulsion from the core. Thus, in the case of 
$B_{c0}=10^{11}, 10^{10}$ and $10^{9}$ G, $B_{res} \approx 10^{10}, 
2 \cdot 10^{9}$ and $1.5 \cdot 10^8$ G, respectively.\\
\noindent It is worth mentioning that the choice of a stiffer EOS leads to
a more extended phase of {\rm forward creeping}. For such 
stiff EOS and relatively strong core fields ($B_{c0}/B_{p0} \ge 0.1$) after 
about $10^8$ years even the {\rm reverse creeping} stage can be entered.\\

\subsection{The effect of different submergence depths}

A possible explanation for the discrepancy between the estimated rates of 
supernovae and of pulsar births in the galaxy (Frail 1998) is the delayed
switch--on of a pulsar. This feature has been discussed by Muslimov \& Page
(1995), who considered a relatively shallow submergence of the initially present
MF. However, the immediate consequence of a supernova (type Ib or II) explosion
will be a fall back of matter on to the new born NS (Colpi, Shapiro \&
Wasserman, 1996). In case this fall back accretion is hypercritical the ram
pressure of the back--falling matter is larger than the pressure of a, say,
$10^{12}$ G initially present MF and the accretion flow is purely
hydrodynamic; conditions as could be estimated for the SN 1987A (see
Chevalier 1989). Applying the time dependence of the fall back accretion rates
given by Colpi et al. (1996), Geppert, Page \& Zannias (1999) calculated the
submergence of a NS MF for different EOS as a function of the total amount of
accreted matter $\Delta M$. It became clear that for values as inferable for 
SN 1987A (Chevalier 1989) the MF has been submerged so deeply that the created 
NS will not shine up as a radiopulsar for more than $10^8$ years.\\
\noindent When the fall back accretion ceases the submerged field will start the
rediffusion towards the surface of the NS. Once the NS is an isolated one during
all its life, for given conductive properties of the layers above the field
maintaining electrical currents and a given EOS determining the cooling and the
scalelength of the field, the submergence density $\rho_{sub}$ down to which the
MF has been buried by fall back accretion is the only parameter which defines
the duration of that rediffusion process $\tau_{red}$. Evidently, the larger 
$\Delta M$ the larger $\rho_{sub}$ and, hence, the larger $\tau_{red}$.\\
\noindent In order to show clearly the effect of the different $\rho_{sub}$ we
will use the same representation as in the preceding section for the medium EOS
with $Q=0.1$. In that way Figure \ 4 shows the evolution of the NS for
$\rho_{sub} = 10^{12}$, and $10^{14}$g cm$^{-3}$, which corresponds
roughly to an total amount of accreted matter of about $3\times 10^{-5}$,
and $ 10^{-2} \ M_{\odot}$, respectively.\\
The process of flux expulsion from a NS with initially submerged MF is
qualitatively different from that of a standard NS. Since at the beginning of
the NS's life, after the fall back has finished, the surface MF $B_{p0}$ is 
practically zero, there is no spin--down and, hence, the velocity of the 
vortices $v_n$ is zero too. Thus, the neutron vortices, because being fixed in 
number and location, will not force the fluxoids to move outward, but will 
impede them to be expelled by the buoyancy force. The immobility of 
the fluxoids and the initial smallness of the crustal forces results in a 
situation where the balance of forces is given only by $F_b + F_n = 0$, where 
the vortex acting force $F_n$ is just compensating the buoyancy, while 
$F_v$ and $F_{crust}$ are many orders of magnitude smaller than $F_n$, $F_b$.\\
\noindent When the field would remain buried there were no chance to expell the
core flux because the number of vortices remains constant in a stationary 
rotating NS. However, depending on $\rho_{sub}$ the submerged crustal field 
rediffuses towards the surface thereby causing an increasing braking of NS's 
rotation. As shown in Figure\ 4 that the spin--down onsets the later
the deeper the crustal field has been submerged. The initially immobile 
neutron vortices start to move outward and the proton fluxoids, tightly bounded 
to them by $F_b$ follow their motion towards the crust, i.e. the fluxoids are 
in the {\rm comoving} state. With increasing $v_p$ the drag force $F_v$ rises 
and, since the crustal force is still negligible, the balance of forces is 
determined by $F_b + F_n + F_v= 0$. During that early epoch the core field does 
not change which results in a constant and positive $F_b$. Since, on the other 
hand both $F_v$ and $F_n$ are negative, the increasing absolute value of drag 
force reduces the absolute value of the vortex acting force. 
By the vertical line with mark $t_1$ we indicated the moment, when
$F_n = 0$;  $t_1$ is $3 \cdot 10^2$ years for $\rho_{sub}=10^{12}$g cm$^{-3}$
and $6 \cdot 10^4$ years for $\rho_{sub}=10^{14}$gcm$^{-3}$. Exactly at that 
moment there is no interaction between the fluxoids and the vortices and the 
balance of forces is given by $F_b +  F_v= 0$.\\
\noindent However, the rediffusion of the crustal field continues beyond
$t=t_1$ and the velocity $v_n$ (and hence $v_p$) increases too. The change in 
the sign of $F_n$ reflects the change in the state of flux expulsion: while 
early on the vortices hampered the movement of the fluxoids, now the positive 
$F_n$ supports the flux expulsion, though it is still $v_n = v_p$.
The process of increasing $F_v$ and $F_n$ proceeds until $F_n$
reaches its maximal value, determined by $F_n=F_n(\omega_{cr})$.
This moment is indicated in Figure \ 4 by $t_2$, which is $6 \cdot 10^2$ years 
for $\rho_{sub}=10^{12}$g cm$^{-3}$ and $7 \cdot 10^4$ years for 
$\rho_{sub}=10^{14}$gcm$^{-3}$. At that point the fluxoids can no longer 
follow the motion of the vortices, which cut now through the fluxoids,
hence, the {\rm forward creeping} regime has been reached.
Close to this moment, the crustal force becomes comparable to $F_b$, however, 
both are small in comparison with $F_n$ and $F_v$. Later on, when the surface 
MF $B_p$ becomes comparable to the core field $B_c$, the transition from 
a ``submerged'' to a ``standard'' field configuration has been performed.
The velocity of fluxoids is no longer determined by $F_n$ and $F_v$,
but instead by $F_{crust} $ and $ F_b$ , a situation which is the same as 
described in KG00 and the evolution goes through the phases of the forward 
creeping, the comoving, and the reverse creeping regime.
It turns out, that the long--term magnetic and rotational evolution does not
depend crucially on $\rho_{sub}$; it affects mainly the NS behaviour in its 
early life and may lead to the phenomenon that a young NS is recognized to be
an old one because its active age, $\tau_a= P/2/\dot P$, is apparently high.
However, especially for very deep submergence of the field, the pulsar may 
also turn to look younger than its real age when the rediffusion progresses
(see Geppert et al 1999). The rediffusion of the initially submerged MF
may lead to some interesting metamorphoses of ages of NSs: the young NSs look 
older or younger than they really are.
For example, from the magnetic evolution of the NS at the right panel of 
Figure 4, one can estimate at $t=130000$ years an active age of about 
$50000$ years, and a braking index $n \equiv 2- \ddot{P}P/{{\dot{P}}^2} 
\approx 1.4$ . The pulsar B1757-24 in SNR G5.4-1.2 has an active age of 
$16000$ years and a braking index $n<1.33$, while its real age  was estimated 
from the morphology of the SNR to be $170000$ years (Gaensler \& Frail 2000). 
Obviously, the assumption of MF submergence in NSs can at least 
partly explain the discrepancy between true and active age, and the occurence 
of a small braking index for this pulsar (Muslimov \& Page 1995).\\
\noindent The other process which generates a strong surface field after the 
NS's birth, the thermoelectric instability, may not account for those apparent 
discrepancies in the ages, because this instability acts effectively at most 
during the first few thousand years of the NS's life.\\ 
\noindent Note however, that with increasing $\rho_{sub}$, $B_{res}$ increases 
while $P_{res}$ decreases. An increase of $\rho_{sub}$ from $10^{12}$g cm$^{-3}$
to $10^{14}$gcm$^{-3}$ causes an increase of $B_{res}$ by a factor of $4$ and
decreases $P_{res}$ by a factor of $2$. This can be understood from the fact
that for larger $\rho_{sub}$ the spin--down starts later. During that longer 
period a small amount of core flux has been expelled and dissipated in the 
inner crust as well as the NS cooled down, thereby enhancing the conductivity 
in the surface layers which decelerates the rediffusion. Thus, in the process 
of rediffusion the surface MF can not reach the same high value as for the 
corresponding case with shallow submergence. Therefore, the spin--down in case 
of the deep submerged MF is less efficient, $P_{res}$ is smaller, which reduces 
the efficiency of the flux expulsion from the core, resulting eventually in a 
larger $B_{res}$.\\

\subsection{Shear stresses induced by flux expulsion}

\begin{figure}
\centering\includegraphics[width=8cm]{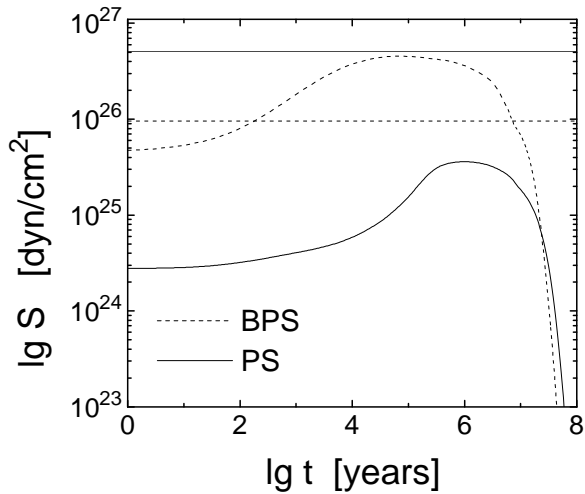}
\caption{Temporal evolution of the shear stresses at the bottom of the
crust generated by the movement of the fluxoids, calculated by use of 
equation\ (\ref{S_crust}) for the soft and the stiff EOS. The maximum  
stresses, given by equation\ (\ref{S_max}) for 
$\mu = 2\cdot 10^{29}$ dyn/cm$^2$ and $\theta_{max} = 10^{-2}$, are shown 
as horizontal lines.}
\end{figure}

\noindent In this paper the crust of the NS is assumed to be solid and not 
deformed by Maxwell (and others) stresses, i.e. the crustal MF evolves only 
through Ohmic dissipation and does not affect the elastic properties of 
the crust. The opposite case was considered by Ruderman (1991a,b) and Ruderman, 
Zhu and Chen (1998), who investigated the consequences of the changing core MF 
configuration for the spin history, especially glitches. Moving the roots of 
the fluxoids at the crust--core interface, crustal strains are built up and 
relaxed by large--scale crust--cracking events. From our calculations we can 
estimate the shear stress at the crust--core boundary. The total force, acting 
upon the crust by the expelled flux is balanced by the force, acting from the 
crust on the fluxoids, i.e. $F_{crust} $. The maximum of $\left| F_{crust} 
\right|$ is reached at that stage of evolution, when the balance of $F_b$ 
and $F_{crust}$ determines the velocity of fluxoids. It seems to be a rather 
common situation, that the maximum force, which acts on the crust, does not 
exceed greatly  the buoyancy force, which acts on the fluxoids from the
beginning of the NS evolution. This force is about $10^{37}$ dyn, as one can 
see from Figure\ 1, for $B_{c0}=2\cdot 10^{12}$ G. The shear stress exerted at 
the base of the crust is given by

\begin{equation}
S_{crust}(t) \sim \frac{ \left| F_{crust}(t) \right|}{R_c \Delta R},
\label{S_crust}
\end{equation}

\noindent where $\Delta R$ is the thickness of the crust and
$\left| F_{crust}(t) \right|$ is evaluated by use of equation
(\ref{F_c_eq}). In Figure\ 5 we compare $S_{crust}$,
calculated for $B_{c0}=2\cdot 10^{12}$G and $Q=0.1$, with the maximum shear
stress the crust can sustain before yielding, as estimated by (Ruderman 1991a),

\begin{equation}
S_{max} \sim \frac{\Delta R}{R}\mu\theta_{max} \la 3 \cdot 10^{26}
{\rm dyn/cm}^2,
\label{S_max} 
\end{equation}

\noindent where $\theta_{max}$ is the elastic strain limit, 
$\theta_{max}\la 10^{-2}$ for a $10^6-10^8$ years old NS, and 
$\mu \la 10^{30}$ dyn cm$^{-2}$ is the shear modulus of the crust.
It is seen that the shear stresses reach their maximum after $10^4-10^6$ years 
and become negligible when the magnetic flux is expelled from the core.
Clearly, the probability for crust--cracking increases with a softening of 
the EOS. Therefore, while for NSs with a soft EOS early crust--cracking events,
as glitches, may occur, their appearance in young ($t < 10^5$ yrs) NSs with 
stiffer EOS hints to an initial field strength larger than $2\cdot 10^{12}$ G 
and/or to a smaller yield strain ($\sim 10^{-4}$) of the young NS crust.\\

\section{Discussion and conclusion}
In order to investigate the magnetic and rotational evolution of isolated NSs
whose MF penetrates both its core and crust, we solved the system of equations
which describe the expulsion of the MF from the superfluid core and its 
subsequent Ohmic dissipation in the solid crust. Thereby we assumed that the
spin--down is driven by magneto--dipole radiation.\\
\noindent In a recent paper (KG00) we considered this process for a NS whose
state of matter is described by a medium EOS and discussed the effects of
different initial field strengths and impurity concentrations in the crust.
For initial field strengths larger than $10^{12}$ G, our study yielded a decay 
time for the surface field  $> 10^7$ years. In the present paper we considered
the effect of different EOS on the MF evolution and its consequences. 
We studied also qualitatively different initial MF structures.\\
\noindent
In the case of a configuration, where the initial MF strengths at the
surface and in the core are of the same order of magnitude (modelled in 
this paper by $B_{p0}=B_{c0}$) we have found, that the main force, which 
drives the fluxoids outward, is the buoyancy force $F_b$. The expulsion 
timescale is determined by the balance of $F_b$ and $F_{crust}$, which depend 
mainly on the core field strength and on the crustal conductivity, and is not 
affected by the spin history of the NS.
The surprising result is that the expulsion timescale depends only weakly on
the EOS. Since stiffer EOS results in a thicker crust, one would expect, that 
$F_{crust}$, which depends on the rate of dissipation of crustal currents, 
counteracts the expulsion of the magnetic flux from the core of a NS with the 
stiffer EOS for a longer time. However, the total core forces are proportional 
to $R_c^3$ (see eq. (\ref{F_eq})), i.e., the total buoyancy force is about 
$6$ times stronger for the PS-model than for the BPS model. The interplay of 
these effects results in rather similar flux expulsion timescales for different 
EOS.\\
\noindent
However, the evolution of the observable surface MF differs substantially
for soft and stiff EOS. For the soft BPS and medium FP (see KG00)
EOS the diffusion timescale of the crustal field  appears to be much shorter 
than the  core flux expulsion timescale. Thus, the evolution of the surface and 
core MFs are tightly connected. Moreover, for the soft and the medium EOS the 
decay timescale of the surface MF depends on its initial strength. In
the case of the thick crust as modelled by the NS with the PS EOS we have 
found, that the diffusion of the crustal field lasts longer than
the expulsion of the core field by approximately two orders of magnitude. That 
means, in case of a stiff EOS the surface field evolution is not sensitive to 
the details of physics of the forces acting upon the fluxoids in the core, and 
does not depend strongly on the initial MF strength. Hence, for a stiff EOS 
the evolution of the surface MF coincides with that, considered by Konar \& 
Bhattacharya (1999), for an initially expelled flux. This result is valid 
for initial surface MFs in the range of $10^{11}-10^{13}$ G, and for impurity 
parameters in the range of $0.01 < Q < 1$.\\
\noindent
It is generally accepted that the MF of normal radiopulsars decays weakly 
during their lifetime (Bhattacharya et al. 1992, Hartman et al. 1996). 
Unfortunately, the results obtained from our scenario of field evolution do not 
allow to select or to reject some EOS as being not in accordance with 
observational facts. All models lead to decay timescales of the surface field
comparable or exceeding the radiopulsars lifetime.\\
\noindent
As in the case of the medium EOS (see KG00), not the entire magnetic flux is
expelled from the core, but some part remains there for eternity,
resulting in a residual MF of NSs. This can be important for the explanation
of the long-living MFs of millisecond pulsars. However, for those NSs a more
complex analysis is required, which takes into account all the effects of 
accretion occuring in binary systems. \\
\noindent
For a ``crustal'' MF configuration, when initially almost the whole magnetic 
flux is confined to the NS's crust, we have found that the neutron vortices 
play a very important role for the dynamics of the fluxoids. The velocity of 
the fluxoids appears to be equal to that of the vortices during almost the 
whole evolution, so that the flux expulsion can be really called
``spin--down induced''(Srinivasan et al. 1990, Jahan Miri \& Bhattacharya 1994).
However, the evolution of the surface MF coincides almost completely with that 
of a purely crustal field as considered by Urpin \& Konenkov (1997),  and Page 
et al. (2000). Correspondingly, all conclusions of those papers are applicable 
to this magnetic configuration, namely the conclusion, that models of NSs, 
based on the medium and stiff EOS with standard cooling yield a satisfactory 
agreement with observations.\\
\noindent For a NS with an initially submerged crustal field we found that the 
submergence affects the evolution only during the early periods of the NS's 
life, as long as the crustal field is diffusing back to the surface. Then, the 
further magnetic and rotational evolution of the NS becomes similar to the 
``standard'' evolution as described in KG00. The only reminder of the 
submergence episode is the relatively weak correlation of the residual field 
strength with the submergence depth and its anti--correlation with the final 
rotational period.\\
\noindent We also calculated the shear stresses at the bottom of the crust, which  arise
when the footpoints of the fluxoids are moved along the crust--core interface.
We found that total force, which acts on the crust from the moved  fluxoids
(being equal to the force, which acts from the crust on to the fluxoids, i.e. 
$F_{crust}$), does not exceed strongly the buoyancy force, which acts onto the 
fluxoids from the  very beginning of the evolution. It can be strong enough to
break  the crust,  especially when a soft EOS decribes the state of the core
matter and  the crust is relatively thin.

\section*{Acknowledgment}

The work of D.K. was supported by a scholarship of the Alexander von
Humboldt--Stiftung and by RFBR grant 00-02-04011. 
We are indebted to the M. Reinhardt and O. Gnedin for carefully reading the
manuscript and for suggestions to improve it.

\section*{References}

\noindent
Baym G., Pethick C., Pines D. 1969, Nature, 224, 673

\noindent
Baym G., Pethick C., Sutherland P.G., 1971, ApJ, 170, 299

\noindent
Bhattacharya D., Wijers R.A.M.J., Hartman J. W., Verbunt F. 1992
A\&A 254, 1990

\noindent
Blandford R., Applegate J., Hernquist L., 1983, MNRAS, 204, 1025

\noindent
Chau H. F., Cheng K. S., Ding K. Y. 1992
ApJ 399, 213

\noindent
Chevalier R. 1989
ApJ 346, 847

\noindent
Colpi M., Shapiro S.L. \& Wasserman I. 1996
ApJ 470, 1075

\noindent
Ding K. Y., Cheng K. S., Chau H. F. 1993
ApJ 408, 167

\noindent
Frail D. A. 1998,
in The Many Faces of Neutron Stars,
eds a. Alpar, R. Buccheri, \& J. van Paradijs,
(Kluwer Academic Press: Dordrecht), p. 179

\noindent
Friedman B., Pandharipande V. 1981, Nucl. Phys. A, 361, 502

\noindent
Gaensler B., Frail D. 2000, 
Nature 406, 158

\noindent
Geppert U., Page D., Zannias T. 1999,
A\&A 345, 847

\noindent
Hartman J., Bhattacharya D., Wijers R., Verbunt F. 1996, A\&A, 322, 477

\noindent
Harvey J., Ruderman M., Shaham J. 1986,
Phys. Rev. D. 33, 2084

\noindent
Itoh N., Hayashi H., Koyama Y. 1993,
ApJ 418, 405

\noindent
Jahan-Miri M., Bhattacharya, D. 1994,
M.N.R.A.S. 269, 455

\noindent
Jahan-Miri M. 2000,
ApJ 532, 514

\noindent
Kluzniak W. 1998,
ApJ 509, L37

\noindent
Konar S., Bhattacharya D. 1999, 
M.N.R.A.S. 308, 795

\noindent
Konenkov D., Geppert U. 2000,
M.N.R.A.S. 313, 66

\noindent
Muslimov A., Page D., 1995, ApJ, 440,L77

\noindent
Muslimov A., Tsygan A. 1985,
Ap\&SS 115, 41

\noindent
Page D. 1998,
in Neutron Stars and Pulsars,
eds N. Shibazaki, N. Kawai, S. Shibata, \& T. Kifune, 
(Universal Academy Press: Tokyo), p. 183.

\noindent
Page D., Geppert U., Zannias T. 2000,
A\&A 360, 1052

\noindent
Pandharipande V. Smith R. 1975,
Nucl. Phys. A. 237, 507
 
\noindent
Ruderman M. 1991a, ApJ, 366, 261

\noindent
Ruderman M. 1991b, ApJ, 382, 576

\noindent
Ruderman M., Zhu T., Chen K. 1998,
ApJ 492, 267

\noindent
Srinivasan G., Bhattacharya D., Muslimov A., Tsygan A., 1990 Curr. Sci., 59, 31

\noindent
Thompson Ch., Duncan R. 1993,
ApJ, 408, 194

\noindent
Urpin V., Konenkov D. 1997,
M.N.R.A.S. 292, 167

\noindent
Urpin V., Levshakov S., Yakovlev D. 1986,
M.N.R.A.S. 219, 703

\noindent
Van Riper, K. 1991,
ApJS 75, 449

\noindent
Wiebicke H.-J., Geppert U. 1996,
A\&A 309, 203

\noindent
Yakovlev D., Urpin V. 1980,
Sov. Astron. 24, 303

\end{document}